\documentclass[aps,twocolumn,showpacs]{revtex4}
\usepackage{amssymb}
\usepackage{graphicx}
\usepackage{bm}
\usepackage{amsmath}
\begin{document}

\title{Oscillatory crossover from two dimensional to three dimensional topological insulators}
\author{Chao-Xing Liu$^{1,5}$, HaiJun Zhang$^2$, Binghai Yan$^3$, Xiao-Liang Qi$^4$, Thomas Frauenheim$^3$, Xi Dai$^2$,
Zhong Fang$^2$ and Shou-Cheng Zhang$^4$}

\affiliation{$^1$ Physikalisches Institut (EP3) and
  Institute for Theoretical Physics and Astrophysics,
  University of W$\ddot{u}$rzburg, 97074 W$\ddot{u}$rzburg, Germany;}

\affiliation{$^2$ Beijing National Laboratory for Condensed Matter
  Physics, and Institute of Physics, Chinese Academy of Sciences,
  Beijing 100190, China;}

\affiliation{$^3$ Bremen Center for Computational Materials Science,
Universit$\ddot{a}$t Bremen, Am Fallturm 1, 28359 Bremen, Germany;}

\affiliation{$^4$ Department of Physics, McCullough Building, Stanford
  University, Stanford, CA 94305-4045;}

\affiliation{$^5$ Center for Advanced Study, Tsinghua
  University,Beijing, 100084, China;}

\date{\today}

\begin{abstract}
We investigate the crossover regime from three dimensional
topological insulators $Bi_2Te_3$ and $Bi_2Se_3$ to two dimensional
topological insulators with quantum spin Hall effect when the layer
thickness is reduced. Using both analytical models and
first-principles calculations, we find that the crossover occurs in
an oscillatory fashion as a function of the layer thickness,
alternating between topologically trivial and non-trivial two
dimensional behavior.
\end{abstract}

 \maketitle

{\it Introduction-} Recent discovery of the two dimensional (2D) and
three dimensional (3D) topological insulator (TI) state has
generated great interests in this new state of topological quantum
matter\cite{bernevig2006d,koenig2007,fu2007a,
hsieh2008,zhang2009,xia2009,chen2009}. In particular, $Bi_2Te_3$ and
$Bi_2Se_3$ are predicted to have bulk energy gaps as large as $0.3
eV$, and gapless surface states consisting of a single Dirac
cone\cite{zhang2009}. Angle-resolved-photo-emission-spectroscopy on
both of these materials observed the single Dirac cone linearly
dispersing from the $\Gamma$ point\cite{xia2009,chen2009}. These
materials have a layered structure consisting of stacked quintuple
layers (QL), with relatively weak coupling between the QLs.
Therefore, it should be relatively easy to prepare these materials
in the form of thin films, either by nano-ribbon growth
method\cite{peng2009}, or by molecular beam epitaxy\cite{li2009}. In
the limit when the thickness $d$ of the thin film are much smaller
than the lateral dimensions of the device, it is natural to ask
whether or not the resulting 2D system is a 2D TI similar to the
$HgTe$ quantum wells\cite{bernevig2006d,koenig2007}. In this letter,
we investigate this question, and find a surprising result that the
crossover from the 3D to the 2D TI occurs in a oscillatory fashion
as a function of the layer thickness $d$.

{\it Effective model analysis-} We begin by recalling the 4-band
effective model of the 3D TI introduced by Zhang {\it et
al}\cite{zhang2009} with the following Hamiltonian:
\begin{eqnarray}
    &&H_{3D}({\bf k})=
    \left(
    \begin{array}{cccc}
        \mathcal{M}({\bf k})&A_1k_z&0&A_2k_-\\
        A_1k_z&-\mathcal{M}({\bf k})&A_2k_-&0\\
        0&A_2k_+&\mathcal{M}({\bf k})&-A_1k_z\\
        A_2k_+&0&-A_1k_z&-\mathcal{M}({\bf k})
    \end{array}
    \right)\nonumber\\
    &&+\epsilon_0({\bf k})
    \label{eq:Heff}
\end{eqnarray}
with $k_\pm=k_x\pm ik_y$, $\epsilon_0({\bf
k})=C+D_1k_z^2+D_2k_\perp^2$ and $\mathcal{M}({\bf
k})=M_0+B_1k_z^2+B_2k_\perp^2$. This Hamiltonian has been
successfully used to discuss the property of V-VI semiconductor,
such as $Bi_2Se_3$ and $Bi_2Te_3$\cite{zhang2009}. The four basis of
the above effective Hamiltonian are denoted as
$\left|P1^+_z,\uparrow\right\rangle,
\left|P2^-_z,\uparrow\right\rangle,
\left|P1^+_z,\downarrow\right\rangle,
\left|P2^-_z,\downarrow\right\rangle$ with the superscript $\pm$
standing for even and odd parity and $\uparrow$ ($\downarrow$) for
spin up (down). An important feature is that the two orbitals $P1_z$
and $P2_z$ have the opposite parities, so that the off-diagonal term
are linear in $k_z$ and $k_\pm$. Another important model is the
4-band effective model proposed by Bernevig, Hughes and Zhang
(BHZ)\cite{bernevig2006d} for 2D quantum spin Hall (QSH) insulator,
given by the effective Hamiltonian
\begin{eqnarray}
    &&H_{2D}({\bf k})=
    \left(
    \begin{array}{cccc}
        \tilde{\mathcal{M}}({\bf k})&0&0&\tilde{A}k_-\\
        0&-\tilde{\mathcal{M}}({\bf k})&\tilde{A}k_-&0\\
        0&\tilde{A}k_+&\tilde{\mathcal{M}}({\bf k})&0\\
        \tilde{A}k_+&0&0&-\tilde{\mathcal{M}}({\bf k})
    \end{array}
    \right)\nonumber\\
    &&+\tilde{\epsilon}_0({\bf k})
    \label{eq:Heff1}
\end{eqnarray}
with $\tilde{\epsilon}_0({\bf k})=\tilde{C}+\tilde{D}_2k_\perp^2$
and $\tilde{\mathcal{M}}({\bf k})=\tilde{M}_0+\tilde{B}_2k_\perp^2$.
The four basis for the $HgTe$ system are taken as
$\left|E1,\frac{1}{2}\right\rangle,
\left|H1,-\frac{3}{2}\right\rangle,
\left|E1,-\frac{1}{2}\right\rangle,
\left|H1,\frac{3}{2}\right\rangle$. According to Ref.
\cite{bernevig2006d}, the orbitals $E1$ and $H1$ also have the
opposite parities, similar to the 3D TI model. The finite size 
effect has been studied previously by B. Zhou {\it et al} 
within the BHZ model.\cite{zhou2008} Another similarity
between these two models is that in order to describe 2D or 3D TI,
we need the condition $M_0B_{1,2}<0$ or $\tilde{M}_0\tilde{B}_2<0$,
so that the system is in the inverted regime\cite{bernevig2006d}. In
fact if we simply take $k_z=0$, the $A_1$, $B_1$ and $D_1$ terms
vanish and 3D TI model (\ref{eq:Heff}) will reduce exactly to the
BHZ model for the 2D TI(\ref{eq:Heff1}). These similarities suggest
that it is possible to give a unified description of both the 2D and
3D TI, and it is helpful to investigate the crossover between them
when the dimension is reduced by quantum confinement. Therefore in
the following we would like to consider the 3D TI model
(\ref{eq:Heff}) in a quantum well or a thin film configuration. For
simplicity, here we assume the film or well thickness to be $d$ and
use an infinite barrier to represent the vacuum.

\begin{figure}
    \begin{center}
        \includegraphics[width=3.3in]{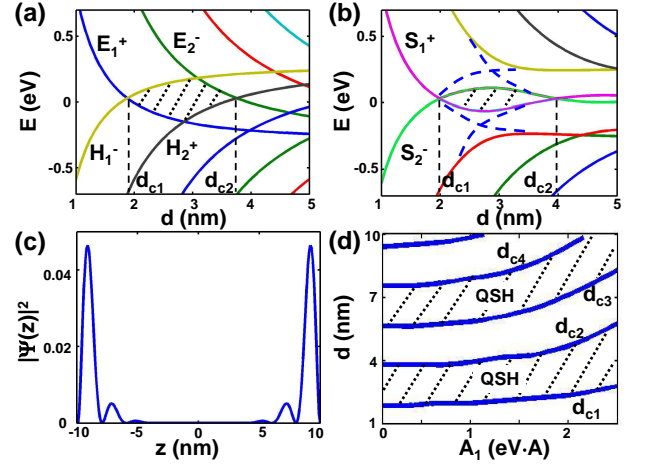}
    \end{center}
    \caption{(Color online) The energy level versus the thickness of
    the quantum well is shown for (a) $A_1=0eV\cdot$\AA, (b)
    $A_1=1.1eV\cdot$\AA. Other parameters are taken from Ref.\cite{zhang2009}.
    The shaded region indicates the regime for QSH states.
    The blue dashed line in (b) shows how the crossing between
    $|E_1(H_1)\rangle$ and $|H_2(E_2)\rangle$ is changed to
    anti-crossing when $A_1$ is non-zero.
    In (c), the density of $|S^{+}_1\rangle$
    ($|S^{-}_2\rangle$ has the same density) is plotted
    for $A_1=1.1eV\cdot$ \AA. In (d), the critical thickness
   $d_{cn}$ ($n=1,2,\cdots$) is
    plotted as a function of $A_1$. QSH states appear in the shaded
    region. }
    \label{fig:band2}
\end{figure}

To establish the connection between 2D BHZ model and 3D TI model, we
start from the special case of $A_1=0$ and turn on $A_1$ later. For
$A_1=0$ the eigenvalue problem of the infinite quantum well can be
easily solved at $\Gamma$ point ($k_x=k_y=0$). The eigenstate are
simply given by
$|E_n(H_n)\rangle=\sqrt{\frac{2}{d}}\sin\left(\frac{n\pi z}{d}
+\frac{n\pi}{2}\right)|\Lambda\rangle$, with $|\Lambda\rangle=
|P1^+_z,\uparrow(\downarrow)\rangle$ for electron sub-bands and
$|\Lambda\rangle=|P2^-_z,\uparrow(\downarrow)\rangle$ for hole
sub-bands, and the corresponding energy spectrum are
$E_e(n)=C+M_0+(D_1+B_1)\left( \frac{n\pi}{d} \right)^2$ and
$E_h(n)=C-M_0+(D_1-B_1)\left( \frac{n\pi}{d} \right)^2$
respectively. Here we assume $M_0<0$ and $B_1>0$ so that the system
stays in the inverted regime\cite{bernevig2006d}. The energy
spectrum is shown in Fig \ref{fig:band2} (a). When the width $d$ is
small enough, electron sub-bands $E_n$ have higher energy than the
hole sub-bands $H_n$ due to the quantum confinement effect. Because
the bulk band is inverted at $\Gamma$ point ($M_0<0$), with
increasing $d$ the energy of the electron sub-bands will decrease
towards their bulk value $M_0<0$, while that of the hole sub-bands
will increase towards $-M_0>0$. Thus there must be crossing points
between the electron and hole sub-bands.
%Due to the inverted
%band structure of bulk materials, at the large thickness $d$ all the
%electron sub-bands have lower energy than the hole sub-bands.
%However when $d$ is reduced, because of the quantum confinement
%effect, the electron sub-bands will increase their energy while the
%hole sub-bands will decrease their energy. Therefore, there must
%exist crossing points between each electron sub-band and hole sub-band.

Now let us focus on the crossing points between the $n$th electron
sub-band $|E_n\rangle$ and the $n$th hole sub-band $|H_n\rangle$,
which occurs at the critical thickness $d_{c1},d_{c2},\cdots$ in
Fig.\ref{fig:band2} (a). Near one of these critical values of $d$,
the low-energy physics can be obtained by projecting the 3D TI model
(\ref{eq:Heff}) into the basis $|E_n,\uparrow(\downarrow)\rangle$
and $|H_n,\uparrow(\downarrow)\rangle$. The resulting effective
Hamiltonian is nothing but BHZ model
%by redefining parameters properly\cite{bernevig2006b}.
(\ref{eq:Heff1}) with $\tilde{C}=C+D_1(n\pi/d)^2$,
$\tilde{M}_0=M_0+B_1(n\pi/d)^2$, $\tilde{A}=A_2$, $\tilde{B}_2=B_2$
and $\tilde{D}_2=D_2$. Following the argument of
BHZ\cite{bernevig2006d}, we know that there is a topological phase
transition between QSH state and ordinary insulator across the
critical point where the gap is closed. The above result can also be
obtained from the parities of the sub-bands $|E_n\rangle$ and
$|H_n\rangle$ at $\Gamma$ point. Since now the wave function has
four components, the parity should be determined by both the
function $\sin\left(\frac{n\pi z}{d}+\frac{n\pi}{2}\right)$ and the
basis $|\Lambda\rangle$. Define $\hat{P}$ as the inversion operator
and we find that $\hat{P}|E_n\rangle=(-1)^{n+1}|E_n\rangle$ and
$\hat{P}|H_n\rangle=(-1)^n|H_n\rangle$, hence $|E_n\rangle$ and
$|H_n\rangle$ always have the opposite parities. According to the
parity criterion of Fu and Kane\cite{fu2007a}, the topological
property of the system can be determined by the product of the
parities of all the occupied energy levels at all time reversal
invariant momenta, which is denoted as $\nu$. Near the critical
thickness $d_{cn}$, the band gap is determined by $|E_n\rangle$ and
$|H_n\rangle$ at $\Gamma$ point and all the other energy levels are
far from the Fermi surface. Therefore, due to their opposite
parities, the crossing of $|E_n\rangle$ and $|H_n\rangle$ will
change the total parity $\nu$ and correspondingly the topological
property of the system. The crossing points can be determined by the
condition $E_e(n)=E_h(n)$, which leads to
$d_{cn}=n\pi\sqrt{\frac{B_1}{|M_0|}}$. A topological phase
transition occurs at $d_{cn}$ for each $n$, so that the system
oscillates between QSH phase and ordinary insulator phase with the
period of $l_c=\pi\sqrt{\frac{B_1}{|M_0|}}$, if the Fermi energy
always stays within the gap.

Next we consider the effect of turning on $A_1$ term. As shown in
Fig. \ref{fig:band2} (b), $A_1$ term induces the coupling between
$|E_n(H_n)\rangle$ and $|H_{n\pm1}(E_{n\pm1})\rangle$, which results
in an avoid crossing between these sub-bands. However, a finite
$A_1$ does not break parity, so that the crossing at $d_{cn}$
between $|E_n\rangle$ and $|H_n\rangle$ with opposite parity remains
robust. Since the topological phase transition is mainly determined
by the crossing between $|E_n\rangle$ and $|H_n\rangle$, the system
still oscillates between QSH insulator and ordinary insulator when
the thickness $d$ is tuned. Due to the hybridization induced by
$A_1$ term, two special states (each doubly degenerate due to spin),
denoted as $|S^+_1\rangle$ and $|S^-_2\rangle$ in
Fig.\ref{fig:band2} (b), are formed within the bulk gap and
well-separated in energy from other sub-bands when the thickness $d$
is large. $|S^+_1\rangle$ is in fact the superposition of
$|E_{2n-1}\rangle$ and $|H_{2n}\rangle$ ($n=1,2,\cdots$) while
$|S^-_1\rangle$ is the superposition of $|H_{2n-1}\rangle$ and
$|E_{2n}\rangle$ ($n=1,2,\cdots$). In large $d$ limit, the gap
between $|S^+_1\rangle$ and $|S^-_2\rangle$ goes to zero and the two
states become nearly degenerate. %Since the quantum well consists of
%TI material, there exist one surface state with gapless Dirac
%dispersion at one surface\cite{zhang2009}.
In Fig.\ref{fig:band2} (c), the density $|\Psi(z)|^2$ for
$|S^+_1\rangle$ or $|S^-_2\rangle$ is plotted as a function of the
position $z$ for a quantum well with the width $d=20nm$, which
indicates that $|S^+_1\rangle$ and $|S^-_2\rangle$ are localized on
the two surfaces of the thin film. One should recall that the bulk
material is a 3D TI with surface states on each
surface\cite{zhang2009}, so that in the large $d$ limit
$|S^+_1\rangle$ and $|S^-_2\rangle$ are nothing but the bonding and
the anti-bonding state of the two surface states at the two opposite
surfaces. Since $|S^+_1\rangle$ and $|S^-_2\rangle$ have opposite
parities, the topological phase transition can also be understood as
the change of sequence between these two states. On top of the
exponential decay, the wave function shown in Fig. \ref{fig:band2}
(c) also oscillates with a period $\sim\pi\sqrt{\frac{B_1}{|M_0|}}$,
which coincides with the oscillation period of the system between
QSH insulator and ordinary insulator. Although $A_1$ term can not
eliminate the crossing between $|E_n\rangle$ and $|H_n\rangle$, it
can shift the critical thickness $d_{cn}$. The dependence for the
critical thickness on the parameter $A_1$ is shown in
Fig.\ref{fig:band2} (d), with the QSH regime labeled by the shading
regions. One can see that when $A_1$ is increased, all $d_{cn}$
($n=1,2,\cdots$) are shifted to the large values, but the
oscillation length doesn't change too much. This indicates that the
QSH regime will always exist, although the gap will be reduced when
$A_1$ is large.

\begin{figure}
    \begin{center}
       \includegraphics[width=3.5in]{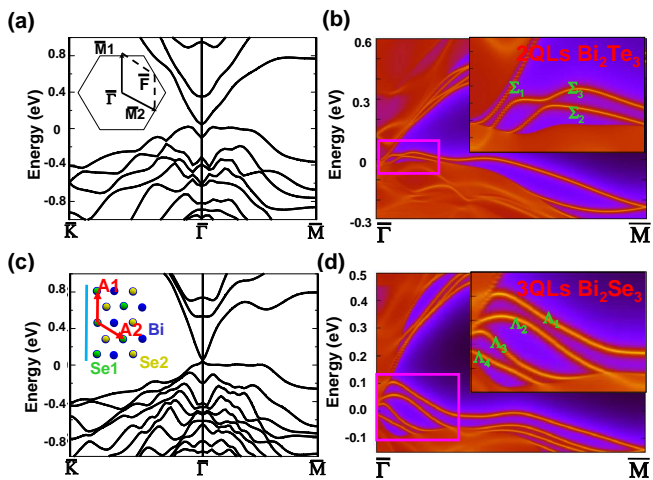}
    \end{center}
    \caption{ (Color Online) Left: the energy dispersion of 2D thin film is
    plotted for (a) 2QLs $Bi_2Te_3$ and (c) 3QLs $Bi_2Se_3$.
    Right: the energy dispersion of the
    semi-infinite film with the edge along A1 direction is plotted for (b)
    2QLs $Bi_2Te_3$ and (d) 3QLs $Bi_2Se_3$. The inset of (a) shows the 2D
    BZ and that of (c) is the top view of 2D thin film with two in-plane
    lattice vectors $A1$ and $A2$. The 1D edge is indicated by the blue
    line along $A1$ direction. The insets of (b) and (d)
    are the zoom-in of the energy dispersion near the bulk gap. }
    \label{fig:edge}
\end{figure}

{\it Realistic materials-} The above discussion based on the
anaytical model gives us a clear physical picture of the crossover
between 2D BHZ model and 3D TI model, and in the following we would
like to consider about the possible realization in the realistic TI
materials $Bi_2Se_3$ and $Bi_2Te_3$. Here we carried out the
first-principles calculations with the BSTATE package\cite{fang2002}
and Vienna ab-initio Simulation Package (VASP)\cite{kresse1996}. The
generalized gradient approximation of PBE-type is used for
exhange-correlation potential and the spin-orbit coupling is taken
into account. For 2D thin film, we construct the free-standing slab
model with crystal parameters taken from the experimental data. For
$Bi_2Se_3$ and $Bi_2Te_3$, five-atom layers, including two Bi layers
and three $Se$ or $Te$ layers, are stacked along z direction,
forming a QL\cite{zhang2009}. There exists strong coupling between
two atomic layers within one QL but much weaker coupling,
predominantly of the van der Waals type, between two QLs. Therefore,
it is natural to regard a QL as a unit for the thin film. We will
first neglect the lattice relaxation effect which depends on the
detail of the lattice environment and address this problem in the
end. As we have discussed above, the topological property can be
determined by the total parity $\nu$, which has been successfully
utilized to predict the 3D TI, such as $Bi_xSb_{1-x}$\cite{fu2007a}
and $Bi_2Se_3$\cite{zhang2009,xia2009}. Here we apply this method to
the thin film with four time-reversal invariant points in 2D
Brillouin zone(BZ), namely $\bar{\Gamma}(0,0)$, $\bar{M}1(\pi,0)$,
$\bar{M}2(0,\pi)$, $\bar{F}(\pi,\pi)$, as shown in the inset of
Fig.\ref{fig:edge} (a). The parity of the wave function at these
points is well-defined and can be easily calculated to determine the
topological property.

\begin{figure}
%    \begin{center}
% \begin{picture}(1,1)
%  \put(10,85){\makebox(0,0)[tr]{\textbf{(a)}}}
%  \put(131,85){\makebox(0,0)[tr]{\textbf{(b)}}}
%    \put(10,0){\makebox(0,0)[tr]{\textbf{(c)}}}
%  \end{picture}
%        \includegraphics[width=2.3in,angle=-90]{parity.eps}
        \includegraphics[width=3.5in]{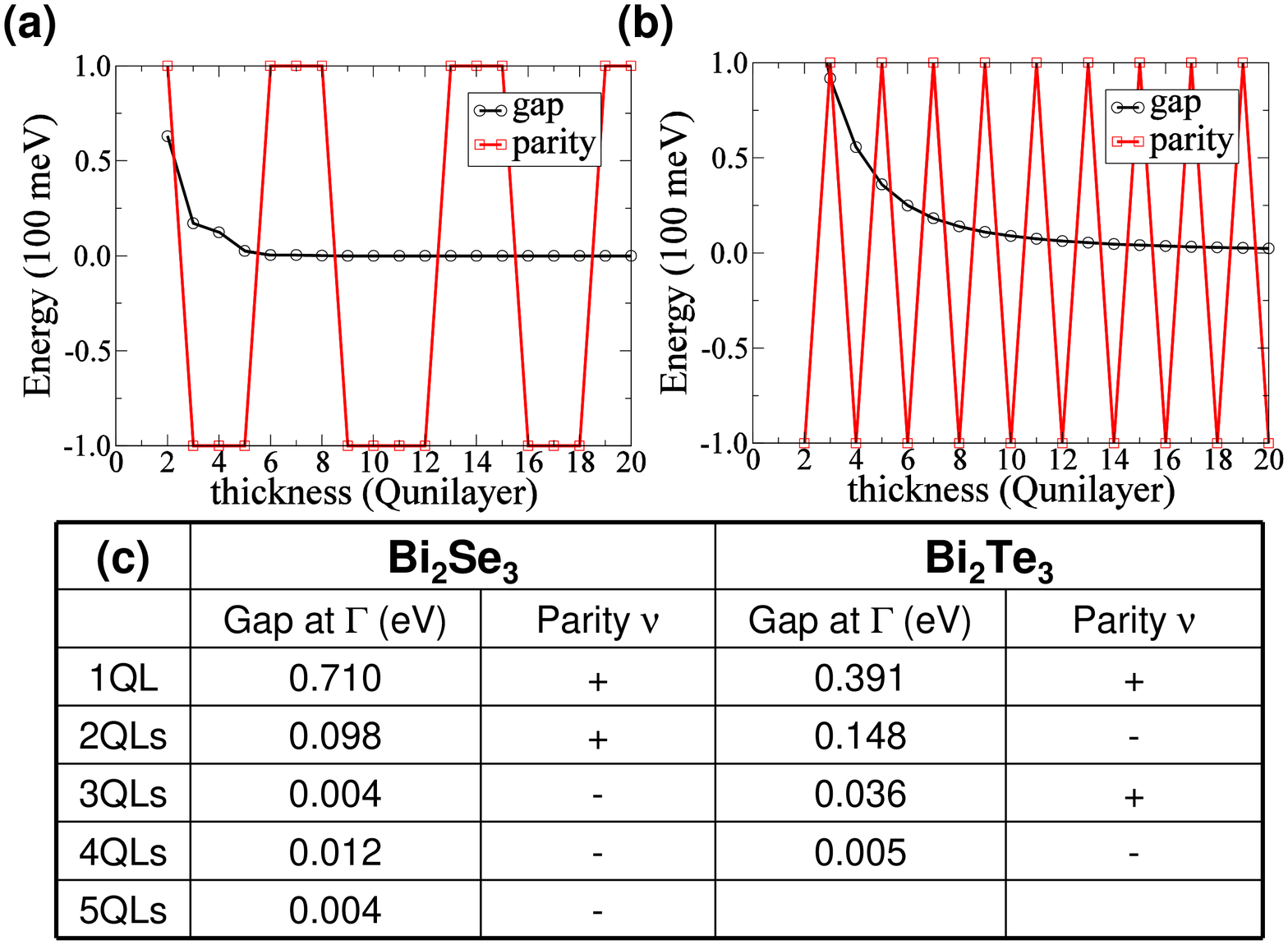}
%        \includegraphics[width=1.6in,angle=0]{gap_BS.eps}
%        \includegraphics[width=1.6in,angle=0]{gap_BT.eps}\\
%        \includegraphics[width=1.1in,angle=-90]{table.eps}
%    \end{center}
    \caption{ (Color online) The band gap and the total parity are plotted as
    the function of the number of the QLs for (a) $Bi_2Se_3$
    and (b) $Bi_2Te_3$. The calculation is based on TB
    model constructed by MLWF from first-principles calculation.
    The results from the fully
    first-principles calculation are shown in table (c) for
    $1\sim 5$ QLs. Good agreement between the TB
    model calculation and first-principles calculation is found. }
    \label{fig:parity}
\end{figure}

The calculated band gap at $\Gamma$ point and the related total
parity $\nu$ for different QLs are shown in the table of
Fig.\ref{fig:parity} (c). The first non-trivial QSH phase appears at
3QLs for $Bi_2Se_3$ and at 2QLs for $Bi_2Te_3$. The gap at $\Gamma$
point of the non-trivial QSH phase for 2QLs $Bi_2Te_3$ is quite
large, of about $148 meV$, however as shown in Fig.\ref{fig:edge}
(a), the $Bi_2Te_3$ is an indirect gap material, and the indirect
gap is an order smaller, only $\sim 10meV$, similiar to the gap of
4QLs $Bi_2Se_3$. When the thickness of the film increases, the time
cost for the {\it ab initio} calculation also increases rapidly,
therefore we perform the tight-binding (TB) model calculation based
on the maximally localized Wannier function
(MLWF)\cite{marzari1997,souza2001} from the {\it ab initio}
calculation, which has also successfully applied to calculate the
surface states of $Bi_2Se_3$ type of materials\cite{zhang2009}. The
band gap and the total parity $\nu$ obtained from the TB calculation
are summarized in Fig. \ref{fig:parity} (a) for $Bi_2Se_3$ and (b)
for $Bi_2Te_3$. The results for the $1\sim 5$ QLs fit with {\it ab
initio} calculation, which confirms the validity of our method. From
Fig.\ref{fig:parity} (a) and (b), it is clear that the system
oscillates between QSH insulator and ordinary insulator,
which verifies our previous analysis of oscillating crossover.

When the 2D system stays in the QSH phase, there are topologically
protected helical edge states at the 1D
edge\cite{kane2005A,wu2006,dai2008}. To show the topological feature
more explicitly, we calculate the dispersion spectra of the helical
edge states directly. As examples, here we study the edge states of
the 2QLs $Bi_2Te_3$ and 3QLs $Bi_2Se_3$ film along $A1$ direction,
as shown in the inset of Fig.\ref{fig:edge} (c). For a semi-infinite
system, combining the TB model from MLWF with the iterative method,
we can calculate the Green's function\cite{dai2008} for the edge
states directly. The local density of states is directly related to
the imaginary part of Green's function, from which we can obtain the
dispersion of the edge states. The topological nature of the edge
states can be determined by the method suggested by Fu and
Kane\cite{fu2007a}. As shown in Fig.\ref{fig:edge} (b) for 2QLs
$Bi_2Te_3$, there exist one edge state $\Sigma_1$ which stays in the
valence band at $\bar{\Gamma}$ point and goes to conduction band at
$\bar{M}$ point. Such edge states are in fact the helical edge
states which can not be eliminated by the local time-reversal
invariant perturbation. For 3QLs $Bi_2Se_3$ in Fig.\ref{fig:edge}
(d), there are three edge states $\Lambda_{1,2,3}$ connecting the
conduction and valence band, which guarantee the system to be
non-trivial. There are also other trivial edge states
($\Sigma_{2,3}$ and $\Lambda_4$) always staying in the conduction
band or valence band, which will not change the topological property
of the system.

In the discussion above, we have only considered the bulk parameters
and have neglected the influence of the surface lattice relaxation.
Since for 2D thin film, the lattice relaxation always plays an
important role in the electronic band structure, we would like to
address this problem in the following. The largest non-trivial gap
appears for 2QLs $Bi_2Te_3$ and 4QLs $Bi_2Se_3$, which is suitable
for experiment to observe. If the thin film is relaxed in vacuum
environment, the non-trivial QSH phase could be changed to ordinary
insulator phase, due to the change of the distance between two QLs.
The reason is that the Van der Waals interaction between two QLs is
very week and quite sensitive to the lattice environment. This also
indicates that the lattice relaxation should depends strongly on the
pressure and the substrate. Therefore, here we propose two different
ways to overcome the negative influence from surface lattice
relaxation. One simple way is to applied uniaxial compressive stress
(along the growth direction) to the film. Based on first-principles
calculation, it is found that only 0.1 GPa stress can recover the
non-trivial gap, which can be easily achieved in
experiment\cite{polvani2001}. Another way is to fabricate the sample
on the proper substrate. For the material with the positive Poisson
ratio, when the lattice is stretched within the film plane, it tends
to get thinner in the perpendicular direction. Therefore, we can use
the materials with the larger in-plane lattice constant than that of
$Bi_2Se_3$ or $Bi_2Te_3$ as the substrate. Since in the present
paper we focus on the principle of realizing QSH insulator from 3D
TI, further study about the effect of pressure and substrate will be
addressed elsewhere.

{\it Conclusion-} We have studied the crossover between 3D and 2D TI
in $Bi_2Se_3$ and $Bi_2Te_3$ systems. Based on both effective model
analysis and {\it ab initio} calculation, we found that it is
possible to obtain 2D QSH state by confining 3D TI in a quantum well
or thin film configuration with the proper thickness. Up to now, 2D
QSH effect has only been observed in $HgTe$ quantum
wells\cite{koenig2007}, and our work may open up a new way to search
for new materials with 2D QSH effect. Recently, the high quality
samples of $Bi_2Se_3$\cite{zhang2009b} and $Bi_2Te_3$\cite{li2009}
thin films have already been fabricated in experiment, so that it is
likely to observe the proposed phenomenon soon.

We would like to thank L.W. Molenkamp for the helpful discussion.
This work is supported by the NSF of China, 
the National Basic Research Program of China (No. 2007CB925000), 
the International Science and Technology Cooperation
Program of China and by the US Department of Energy, Office of Basic
Energy Sciences, Division of Materials Sciences and Engineering,
under contract DE-AC02-76SF00515. 
C.X. Liu and B.H. Yan acknowledge financial support by the Alexander
von Humboldt Foundation of Germany. 

{\it Note added}: During the course of this work, we have learned about the 
work of H.Z. Lu {\it et al}\cite{lu2009} and J. Linder {\it et al}\cite{linder2009}, 
in which the similar finite size effect has been studied within the 3D TI 
model of Zhang {\it et al}\cite{zhang2009}.

\bibliography{crossover}

\end{document}